\documentclass[%
 reprint,
nofootinbib,
 amsmath,amssymb,
 aps,
]{revtex4-2}

\usepackage{hyperref}
\hypersetup{colorlinks=true,citecolor=blue,linkcolor=blue,filecolor=blue,urlcolor=blue}

\usepackage{graphicx}

\usepackage[utf8x]{inputenc}
\usepackage{lmodern,textcomp}

\usepackage{color}

\newcommand\alesia{\textsc{alesia}}

\begin{document}

\title{Data-Driven Optimisation of Superconducting Magnets at CEA Paris-Saclay}

\author{Damien F. G. Minenna}\email{damien.minenna@cea.fr}
\author{Guillaume Dilasser}
\author{Robin Penavaire}
\author{Valerio Calvelli}
\author{Thibault de Chabannes}
\author{Thibault Lecrevisse}
\author{Thomas Achard}
\author{Jason Le Coz}
\author{Christophe Berriaud}
\author{Benoît Bolzon}
\author{Antomne Caunes}
\author{Phillipe Fazilleau}
\author{Hélène Felice}
\author{Clément Genot}
\author{Antoine Guinet}
\author{Nikola Jerance}
\author{François-Paul Juster}
\author{Thibaut Lemercier}
\author{Gilles Lenoir}
\author{Clément Lorin}
\author{Yann Perron}
\author{Camille Pucheu-Plante}
\author{Étienne Rochepault}
\author{Damien Simon}
\author{Francesco Stacchi}
\author{Michel Segreti}
\author{Vincent Trauchessec}
\author{Olivier Tuske}
\author{Hajar Zgour}

\affiliation{Université Paris-Saclay, CEA, IRFU, Département des Accélérateurs, de la Cryogénie et du Magnétisme, Gif-sur-Yvette, 91191, France}

\date{\today}

\begin{abstract}
Superconducting magnets for particle accelerators are particularly challenging to design because they involve a large number of coupled physical phenomena and the management of complex datasets.
Artificial Intelligence (AI), including machine learning and advanced optimisation techniques, offers promising approaches to address these challenges and accelerate the design process.
This paper presents a new AI-based optimisation and data management platform, and highlights several ongoing applications of AI methods carried out at CEA Paris-Saclay, including multiphysics optimisation using active learning, topology optimisation, holistic modelling of an Electron Cyclotron Resonance (ERC) ion source, and anomaly detection in quench events.
\end{abstract}




\maketitle

\section{Introduction}
\label{s:intro}

Since the 1960s, CEA Paris-Saclay has been actively involved in the design, construction, and operation of superconducting magnets for a wide range of applications including: particle accelerators and detectors, fusion devices, and Magnetic Resonance Imaging (MRI) systems.
Superconducting magnets are highly integrated and complex systems. 
Their development is both multidisciplinary and data-intensive, requiring expertise across multiple areas of physics and engineering, including electromagnetism, mechanics, thermal analysis, cryogenics, and materials science. 
Accurate modelling must account for the various components of the magnet system, such as superconducting cables, winding geometries, ferromagnetic structures, cooling circuits, quench and safety protection systems, and power supplies.
The design and construction of such magnets typically span several years, during which project requirements may evolve.
During the design phases, extensive numerical simulations are performed, often relying on computationally expensive Finite Element Method (FEM) models to capture strongly coupled multiphysics behaviours.
In addition, the emergence of new High-Temperature Superconductors (HTS) requires the development of new modelling and optimisation tools.

In this context, the integration of Artificial Intelligence (AI), including Machine Learning (ML) and advanced optimisation techniques, is transforming the way superconducting magnets are designed and operated \cite{YazdaniAsrami2022}, particularly through three main contributions.
First, AI approaches can significantly reduce computational time and resource requirements through the development of surrogate (reduced-order) models.
Second, AI enables more efficient and systematic optimisation strategies, particularly for identifying optimal configurations in high‑dimensional parameter spaces.
Third, AI methods provide new opportunities to improve operational efficiency, anomaly detection and overall system safety during magnet operation. 
This paper presents several examples related to superconducting magnets that illustrate these three aspects, based on recent activities carried out at CEA.
CEA is also investigating in this paper the use of those tools towards the development of the new Electron Cyclotron Resonance Ion Sources (ECRIS).

This paper is organised as follows: Section~\ref{s:alesia} presents a new platform. Section~\ref{s:magnet} presents applications to magnet optimisation, including active learning and topology optimisation. Section~\ref{s:ecr} presents applications to the future ECRIS, including HTS magnets and the extraction system.
Finally, Section~\ref{s:quench} addresses quench detection and simulation.

\section{Data Analysis And Management}
\label{s:alesia}

\begin{figure}
    \centering
    \includegraphics[width=\columnwidth]{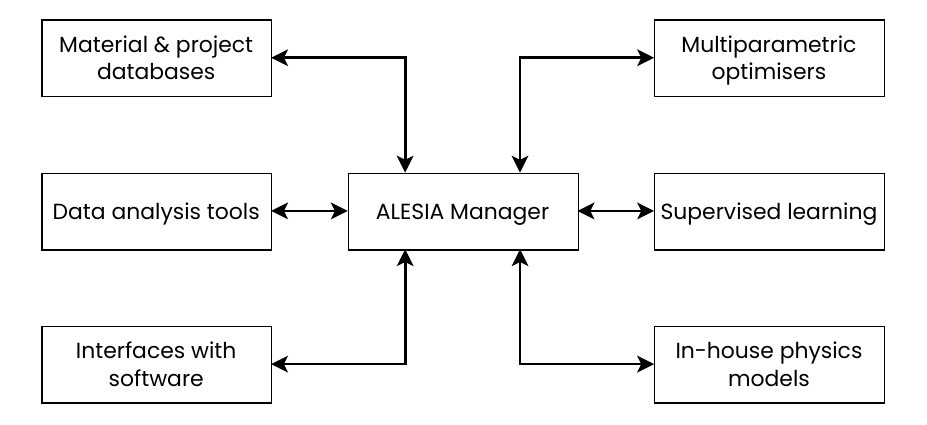}
    \caption{Main modules of the \alesia{} platform.}
    \label{fig:alesia}
\end{figure}

Two of the biggest challenges of designing large superconducting magnets are the large number of parameters involved and the diversity of expertise required across multiple areas of physics and engineering. 
To solve those challenges, CEA is developing a new artificial intelligence optimisation and data management platform called \alesia{}.
\alesia{} is developed in Python and Cython owing to the languages versatility and strong community support, which facilitates the integration of AI tools. 
The platform has been benchmarked on Windows, Linux, and macOS systems, as well as on high-performance computing (HPC) clusters using on job schedulers, such as SLURM.
Note that several other frameworks dedicated to the design of superconducting magnets have been proposed in recent years \cite{Bortot2018, Zani2020, Maciejewski2023}.

It is currently structured into seven main modules as represented in Fig.~\ref{fig:alesia} and described in the following Sections. 

\subsection{Project Manager}

The core of \alesia{} rests on its project manager approach.
A ``project'' is defined as the set of simulation scripts, execution shell scripts, input parameter ranges, individual simulation inputs, optimiser hyperparameters, and the generated data.
The manager is responsible for launching simulations and supports their parallel execution.
It also ensures the coordination and interaction between all modules.

\subsection{Databases}

Such projects can require substantial storage capacity.
For example, one thousand simulations performed with \textsc{opera 2D} \cite{opera} can generate several terabytes of data if full meshed models are stored.
To limit storage requirements, only relevant output data are retained. 
All scripts and input parameters are archived, allowing simulations to be reproduced if necessary. 
The project database is stored in a single file using the HDF5 format.
For each individual simulation, a temporary local HDF5 file is first created and subsequently merged into the project database upon completion. 
This database follows several of the FAIR (Findable, Accessible, Interoperable, Reusable) guiding principles.
It facilitates data analysis by ML models, allows cross-disciplinary uses, and reduces the time spent locating and preparing datasets.

To facilitate design simulations throughout all development phases, a unified material properties database implemented in SQL (Structured Query Language) has also been developed. 
The database stores a wide range of material properties, including magnetisation curves, Young's modulus, shear modulus, Poisson's ratio, thermal expansion coefficients, heat capacity, thermal conductivity, electrical resistivity, density, together with references to the original data sources.
In addition, the database includes critical current density fits for several superconductors, including HTS and Low-Temperature Superconductors (LTS), such as Nb-Ti, Nb$_3$Sn.
A key advantage of this approach is that a consistent material dataset can be accessed by all simulation codes throughout the design workflow (via the interfaces introduced in Section~\ref{s:interface}), thereby limiting the risk of modelling errors. 
A Django-based API (Application Programming Interface) was made to access the material properties database within the CEA network.

\subsection{Interface With Physics Software}
\label{s:interface}

The platform integrates seamlessly with external physics simulation codes. 
This integration serves two primary objectives: exporting input data from \alesia{} to the simulation code and importing the resulting output data back into the \alesia{} project database.

At present, \alesia{} is interfaced with \textsc{opera} \textsc{2d} and \textsc{3d}, \textsc{rat} (open source version) \cite{rat}, \textsc{ansys} \cite{ansys}, \textsc{cast3m} \cite{castem} and \textsc{ibsimu} \cite{Kalvas2010}. Interfaces also exist for codes written in C++, Matlab and Python. 
An interface with \textsc{comsol} \cite{comsol} is currently under development.

Two complementary approaches are employed, sometimes in combination: 
i) a direct database interaction, in which the physics codes are modified to interface directly with the \alesia{} project database. 
As an example, the Python interface in recent versions of \textsc{opera 2D} is exploited via a dedicated package developed for this purpose. 
Similarly, tailored processes have been implemented for \textsc{cast3m}.
ii) a keyword-based script modification, in which input scripts are automatically edited by replacing predefined keywords with the desired values. 
A third-party post-processing script is then used to retrieve the output and add them into the project database. 
This approach is useful when direct modifications of HDF5 files is impractical or unsupported.

\subsection{Multiparametric Optimisers}
\label{s:optimisermethods}

\alesia{} is capable of solving nonlinear optimisation problems with multiple objective functions subject to a set of constraints. 
Numerous heuristic multi-parameter optimisation methods have been implemented and benchmarked in Ref.~\cite{Minenna2026}. They are based on evolutionary computation and active learning, including genetic algorithms, differential evolution, particle swarm optimisation, Covariance Matrix Adaptation Evolution Strategy (CMA-ES), and Bayesian optimisation.
The benchmark was performed on an Nb-Ti superconducting magnet design that simultaneously involved electromagnetism, mechanics, materials, conductor behaviour, and quench simulations. 
The main challenge addressed was managing the highly coupled interdependencies between numerous design parameters.

Previously, in a similar vein, genetic algorithms were employed for the development of the FCC Flared-end Dipole Demonstrator (F2D2) project \cite{Felice2019}, within a collaboration between CEA and CERN, as well as for the MADMAX (MAgnetized Disk and Mirror AXion) experiment \cite{Calvelli2020}, hosted at HERA, DESY.

In addition, several custom optimisation algorithms with different objectives have been developed.
For instance, this includes a parameter sweep over a uniform N-dimensional grid, where each sampled point is further refined using a bisection method. 
This approach is useful when investigating variations in the winding geometry, while requiring the current density to be adjusted at each point to maintain the target magnetic field.
It also includes a Monte Carlo approach for error analysis, in which the parameters of a validated design are randomly perturbed to assess tolerances.

\subsection{Supervised Learning Techniques}

The multi-parametric optimisation algorithms, described in Section~\ref{s:optimisermethods}, are not the only means of exploring designs within \alesia{}.
Once a sufficiently large dataset has been generated (preferably through random sampling), the platform can exploit a variety of supervised learning techniques to construct surrogate models representing the behaviour of the magnet. 
These surrogate models are valuable when, for instance, target requirements are modified, as they enable the rapid identification of new feasible regions within the parametric design space without the need for running extensive additional simulations.

\subsection{Data Analysis Tools}

Data analysis is performed mainly in Python, using customised Jupyter notebooks. 
Those tools allow rapid exploration of the parametric space and facilitate the validation of design results.
A metadata database is maintained throughout the simulations, storing, for each input and output quantity, its description, physical units, symbolic name, and \LaTeX{} representation for use in plots and reports. 
Based on the simulation results, \alesia{} can automatically generate basic reports in \texttt{.docx}, \texttt{.pptx}, or \texttt{.tex} formats.

\subsection{In-House Modelling}

A set of analytical and quasi-analytical models for superconducting magnets are also directly included into \alesia{}, for example Biot-Savart field modelling.
Including those models into \alesia{} allows them to directly access the material database and superconductor property fits.
This capability is particularly useful for the conductors design model and the quench model implemented in the platform.
The conductors design model estimates the conductor composition and margins.
The quench model is based on a 3D adiabatic expansion \cite{Wilson2002} of the normal zone temperature combined with a matrix-based coupled equivalent circuit, allowing to compute different quantities, such as hotspot temperature, current decay and voltages at the terminals of the coils. It is based on the QTRANSIT code developed by Lesmond and Fazilleau \cite{Fazilleau2006, Fazilleau2010}.

\section{Magnet Optimisation Design}
\label{s:magnet}

\subsection{Multiphysics Optimisation}
\label{s:nb3sn}

\begin{figure}
    \centering
    \includegraphics[width=\columnwidth]{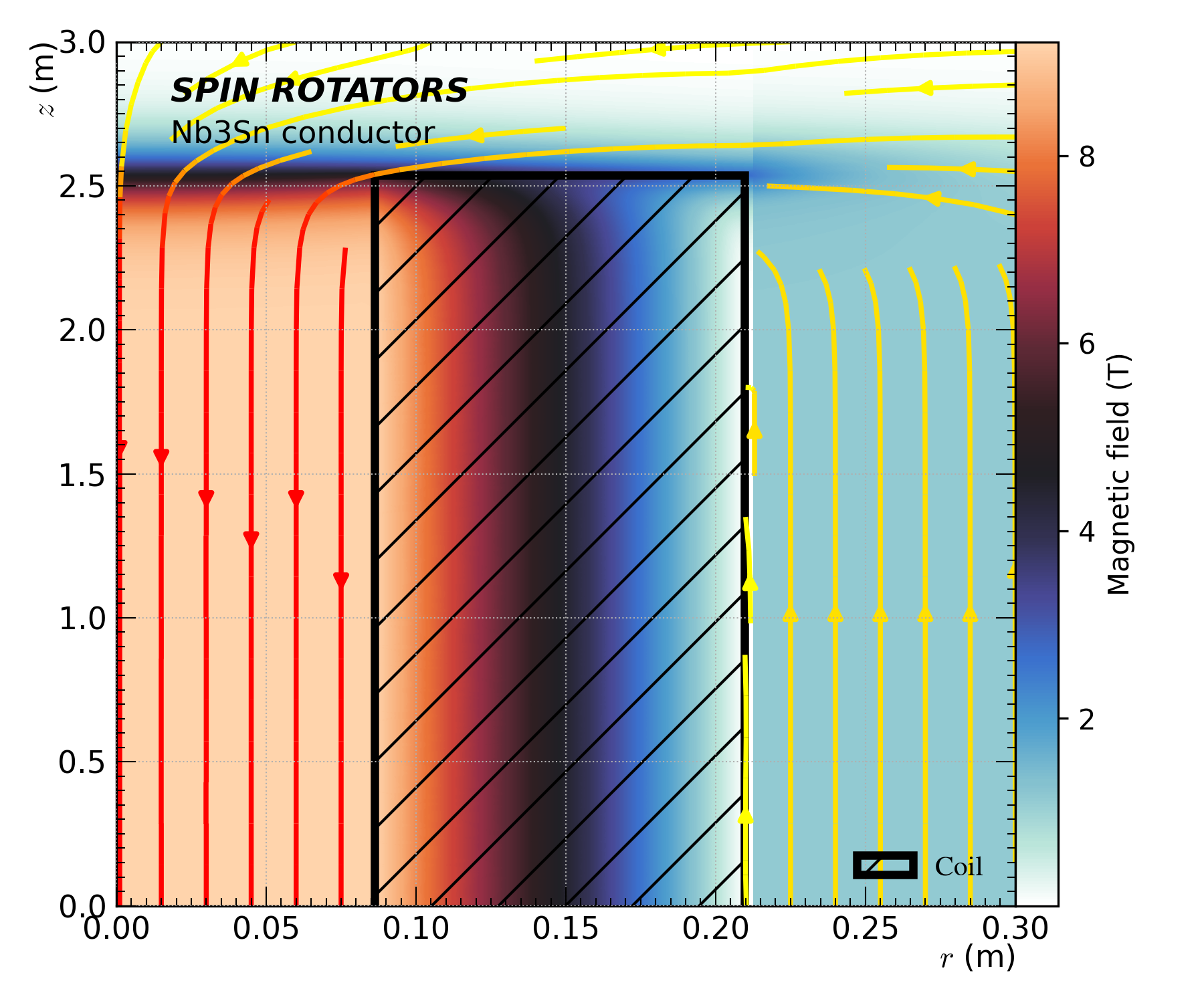}
    \caption{Cross section of the 6.2 m-long SPIN ROTATORS magnet. The colormap represents the magnetic field distribution within the coil. The design was obtained using the \alesia{} platform.}
    \label{fig:spinshort}
\end{figure}

\begin{figure}
    \centering
    \includegraphics[width=\columnwidth]{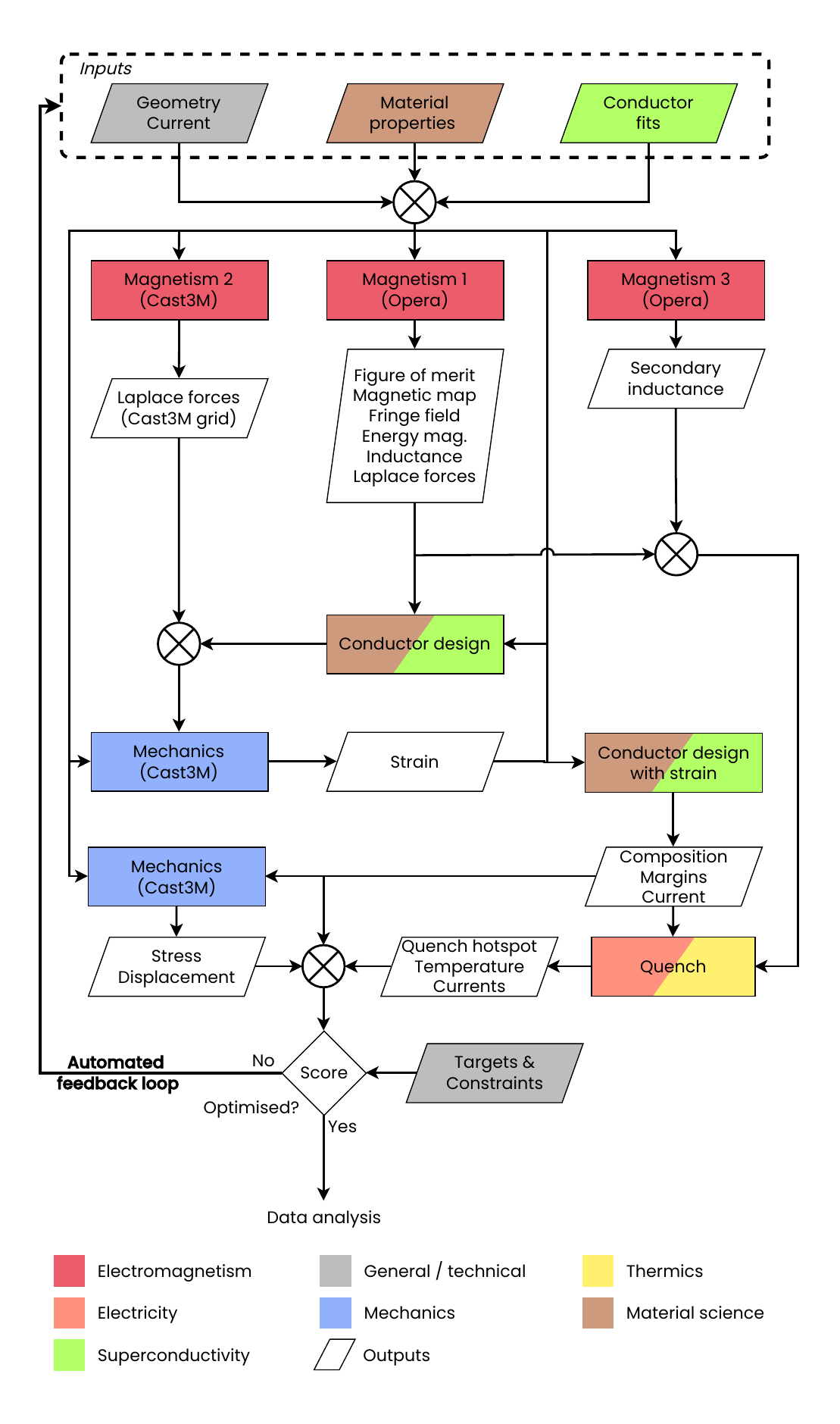}
    \caption{Automated workflow of the design of an Nb$_3$Sn solenoid magnet with the new platform \alesia{}.}
    \label{fig:loop}
\end{figure}

CEA is currently responsible for the design of the SPIN ROTATORS superconducting magnets, which consist of eight solenoid magnets for the Electron-Ion Collider (EIC) \cite{Willeke2021} at Brookhaven National Laboratory (BNL). 
These magnets will control the electron spin polarisation prior to collision, and they require both a high operative magnetic field (around 8~T) and a large magnetic field integral (15.43 T$\cdot$m for the 2.5 m-long magnet and 47.26 T$\cdot$m for the 6.2 m-long magnet , see Fig.~\ref{fig:spinshort}).
To meet these requirements, two different superconductors: Nb-Ti and Nb$_3$Sn were considered. 
The optimisation of both was automated by the \alesia{} platform.
The full design process must integrate magnetic, mechanical, conductor, and quench simulations across multiple materials and variable geometries, including the coil, yoke, mandrel, and insulation layers.

Figure~\ref{fig:loop} displays the design workflow for an Nb$_3$Sn solenoid magnet, illustrating the complexity and interdependence of the different physical domains and associated simulations.
At the beginning of the loop, the values of the design variables (design inputs) are selected by the optimisation algorithm (evolutionary computation or active learning).
A first magnetic simulation is then performed using \textsc{opera 2d} to retrieve the magnetic properties. 
A second magnetic simulation with \textsc{cast3m} is carried out to compute the Lorentz forces.
In addition, a third magnetic simulation with \textsc{opera 2d} evaluates the secondary inductance induced in the stainless steel mandrel.
Next, an initial conductor design estimation is performed using an in-house software to determine a preliminary composition, which is then as input for a first mechanical simulation with \textsc{cast3m}, to evaluate the strain experienced by the Nb$_3$Sn cable. 
Based on these results, the final conductor design is produced, followed by the final mechanical simulation.
A quench simulation ends the workflow.  
Thanks to \alesia{}, the overall process is linear and automated. 
The platform sequentially executes each simulation software with the required inputs, and retrieves the outputs and stores them to the next stage of the workflow. 
All the generated data are stored for subsequent analysis and design verification.
This architecture requires only $N$ interfaces, instead of $(N-1)!$ if direct interfaces had to be implemented between every pair of software.
The complete Nb$_3$Sn workflow is described by 70 parameters, including geometrical variables, current, operating temperature, material and superconductor properties interpolated at the relevant field and temperature.
A the end of the workflow, each design is evaluated through a fitness score, defined as a weighted difference between target and obtained values. 
The optimisation algorithm goal is to minimise this score.
This approach ensures a rapid and comprehensive evaluation of each design, while enabling the exploration of multiple candidate solutions. 
For instance, several magnet and conductor designs can be assessed, each presenting different trade‑offs such as cost, performance, or protection margins.
Using this methodology, several feasible Nb$_3$Sn conductor solutions have been identified. 
Nb-Ti conductors are also being investigated with a similar workflow \cite{Minenna2026}.

\begin{figure}
    \centering
    \includegraphics[width=\columnwidth]{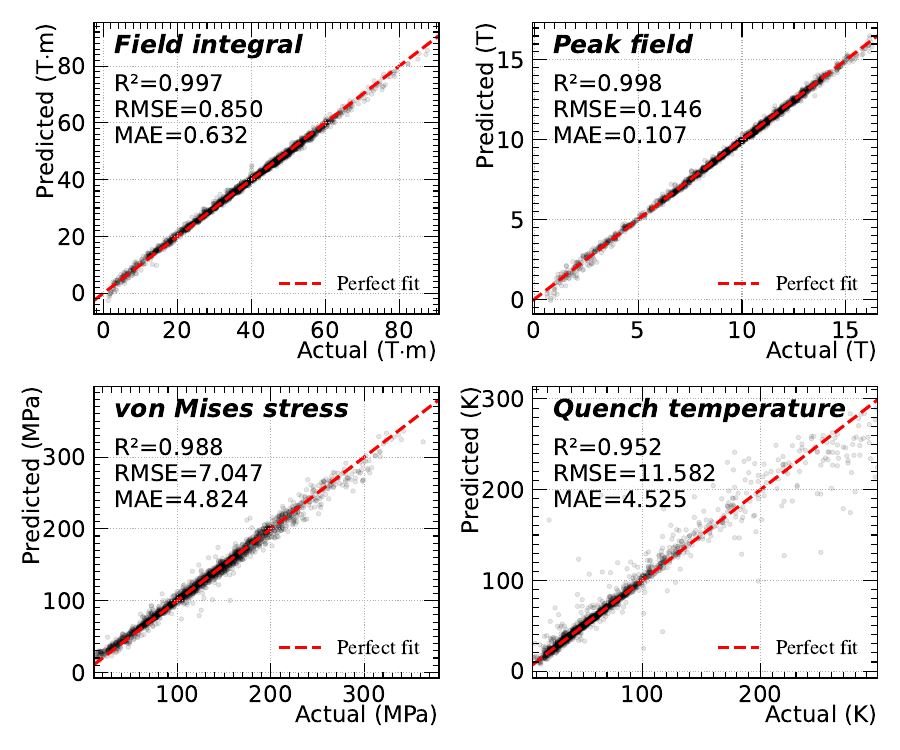}
    \caption{Parity plots with 2131 test points (not used during model training), evaluating a surrogate model predictions. Upper left: integrated magnetic field along the beam propagation axis (reflects beam propagation needs). Upper right: peak magnetic field in the coil (up to 18 T with an Nb$_3$Sn conductor). Lower left: Maximum von Mises stress in the coil windings (relevant to mechanical integrity under Lorentz forces and cooling down). Lower right: Maximum hotspot temperature during a quench event.}
    \label{fig:nb3sn_nn}
\end{figure}

In addition, the data generated from the workflow shown in Fig.~\ref{fig:loop} can be leveraged to train fast surrogate models enabling rapid exploration of the parameter space.
Figure \ref{fig:nb3sn_nn} presents the parity plots obtained from a deep residual neural network designed for multi-output regression, implemented using the PyTorch framework.
The model handles mixed-type input (categorical and continuous variables) and was trained on 8524 simulations managed through the project database.
The network architecture consists of a series of residual blocks with batch normalisation, GELU (Gaussian Error Linear Unit) activation functions, and dropout regularisation. Training is performed on GPU (Graphics Processing Unit) using the AdamW optimiser.

\subsection{Shim Coils}

\begin{figure}
    \centering
    \includegraphics[width=\columnwidth]{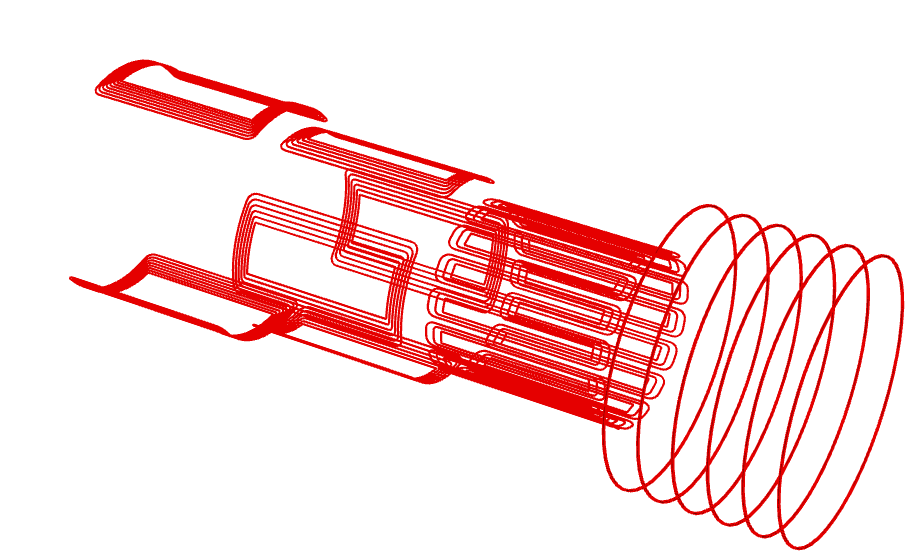}
    \caption{Example of shim coils (racetrack and solenoid geometries). For clarity, the coil positions are shifted along the longitudinal axis. The detector magnet is not represented.
    The system is composed, from left to right, by a dipole (with 8 loops), a quadrupole (with 6 loops), an hexadecapole (with 3 loops) and 6 solenoids.}
    \label{fig:shim}
\end{figure}

Many detectors or MRI magnets require additional shim coils for active shielding and for improving magnetic field homogeneity. 
The shim coil geometry, including multipole order, number of loop per multipole, current density in each loop, length, radius, conductor thickness and width, defines a large parametric space that is explored automatically using \alesia{} Bayesian optimisation.
The optimisation aims at reducing fringe fields and, if needed, can improve local magnetic field homogeneity within the region of interest.
Figure~\ref{fig:shim} shows an example of shim coil systems with a dipole, a quadrupole, an hexadecapole and 6 solenoids for a generic study.
The magnetic field simulations are performed using \textsc{opera 3d}, allowing evaluation of the field distribution produced by complex coil geometries.
In addition, CEA also investigates the use of reinforcement learning methods in order to redesign the shim coil system to innovative configurations beyond traditional design approaches.

\subsection{Topology Optimisation}

\begin{figure}
    \centering
    \includegraphics[width=\columnwidth]{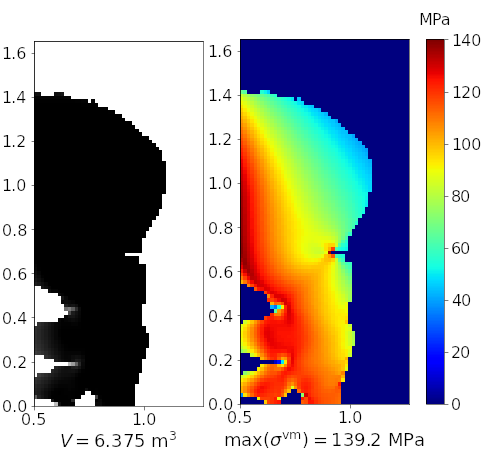}
    \caption{Upper half of the winding cross-section obtained through magneto-mechanical topology optimisation for a 11.7 T MRI solenoid with a Ø1 m bore and maximum von Mises stress below 160 MPa. Left plot: Density map of the optimised solution. White mean empty space, black is windings carrying $j_E = 26.5$ A/mm$^2$. Right plot: Map of von Mises stress in the optimised windings.}
    \label{fig:to}
\end{figure}

Several applications require superconducting magnets capable of generating increasingly high magnetic fields. 
However, higher fields also lead to increased Lorentz forces acting on the coils.  
These forces induce mechanical stresses and deformations that may degrade the superconducting properties of the conductors, and cause irreversible plastic deformation of the windings, or even result in structural failure.  
Consequently, the mechanical integrity of the windings must be considered from the earliest stages of the design process for magnets intended to operate at very high fields.

Another technique currently being investigated to simultaneously handle target field requirements together with mechanical constraints in electromagnet design is based on Topology Optimisation (TO) \cite{Bendsoee1988}.
The methodology is derived from the Solid Isotropic Microstructure with Penalisation (SIMP) approach \cite{Rion2006}. It aims at finding the optimal distribution of the density of homogenised windings inside a predefined discretised design domain. The cost function to be minimised is the volume of the magnet. A mixture of magnetic constraints on the target field and mechanical constraint on the maximum value of the von Mises stress inside the windings are simultaneously considered. The resulting non-linear program is solved using gradient-based optimisers like the Method of Moving Asymptotes (MMA) \cite{Svanberg1987}.

This TO-based approach is presently being developed in 2D for the optimisation of the cross-section of ultra-high field MRI electromagnets \cite{LeCoz2026}. Figure~\ref{fig:to} shows example results of the optimisation process for the cross-section of a highly homogeneous 11.7 T MRI magnet with a Ø1 m bore and a maximum von Mises stress in the windings lower than 160 MPa. The main benefits of this approach are that it takes no initial assumption of the solution topology and that it is very fast, requiring only a few minutes to yield solution on a personal computer. As such, it is believed it could be useful to quickly establish magnet first drafts, that can then be parameterised and refined using the other optimisation approaches presented in this article.

\subsection{HTS optimisation}

As part of CERN's High Field Magnet (HFM) program, CEA is contributing by investigating the possibility of using innovative HTS-based windings, using REBCO (Rare Earth Barium Copper Oxide) tapes, for dipole magnets targeting very high magnetic field. 
Optimising the use of HTS in these magnets requires multiphysics and multiscale approaches, encompassing geometric aspects (minimising hardway bending and the volume of HTS material, optimising the conductor orientation to reduce the anisotropic limitations), electromagnetic aspects (magnetic field quality, minimisation of cylindrical harmonics, generation of magnetic fields exceeding 14 T), electrical aspects (electrical joints, current margins at the cable level, tolerances to local defects, stability and protection against quenches, AC losses), mechanical aspects (Lorentz forces, degradation of superconducting performance due to deformation, changes in contact resistance with pressure), and cooling performances (localisation of thermal drain/channel in relation with instabilities, efficiency of local and global heat transfers). 

In this context, the use of advance optimisation methods within \alesia{}, coupled with an in-house Python wrapper of the \textsc{rat} software (denoted \textsc{pyrat}), enables automated exploration of the design space and identification of the most suitable solutions (see Section~\ref{s:alesia}). 
This approach provides significant time savings by first assessing strong and weak constraints, identifying correlations between parameters, and estimating the range of viable solutions, without the need to exhaustively compute every configuration, which would otherwise requires substantial numerical resources.

\subsection{Beyond particle accelerator magnets}

With the development of particle accelerators, it is possible to achieve a technological transfer toward other application domains. 

In the context of the FASUM (Forty Teslas All Superconducting User Magnet) project, carried out in collaboration with CNRS LNCMI, and continuation of the 32.5 T NOUGAT magnet \cite{Fazilleau2020}, CEA aims to develop a fully user superconducting 40 T user magnet, based on a 19~T LTS ouster, produced by Quantum Design Oxford, and a 21~T HTS insert developed jointly by CEA and CNRS.
To mechanically design the HTS insert, composed of stacked double HTS pancakes, CEA is developing a mechanical analysis procedure based on to numerical Matlab and \textsc{cast3m} codes, enabling the evaluation of stress and strain distributions within the layers induced by the three main loading conditions: pre-tensioning, cooling and energising.
Several challenges remain, especially the distribution of screening currents, which induces  significant stress and strain concentrations. 
In that context, parametric studies using \alesia{} are investigated to identify HTS insert configurations that reduce stress concentrations in critical regions.

In parallel, CEA is also conducting the development of future high-field LTS and HTS MRI magnets. 
The \alesia{} platform is currently deployed for the design of an 11.7~T MRI magnet similar to ISEULT \cite{Quettier2020} using Bayesian optimisation approaches.

\section{ECRIS Design}
\label{s:ecr}

CEA is also heavily involved in the development and operation of ECRIS \cite{Bolzon2025,Tuske2025}.
It is anticipated that the \alesia{} platform will be capable of optimising all aspects of an ECRIS together, from the magnets and plasma, up to the beam extraction and transport. 
For the moment, it is being evaluated on the design of LTS and HTS magnets for ECRIS and their extraction systems.

\subsection{Magnet for next generation ECRIS}

\begin{figure}
    \centering
    \includegraphics[width=\columnwidth]{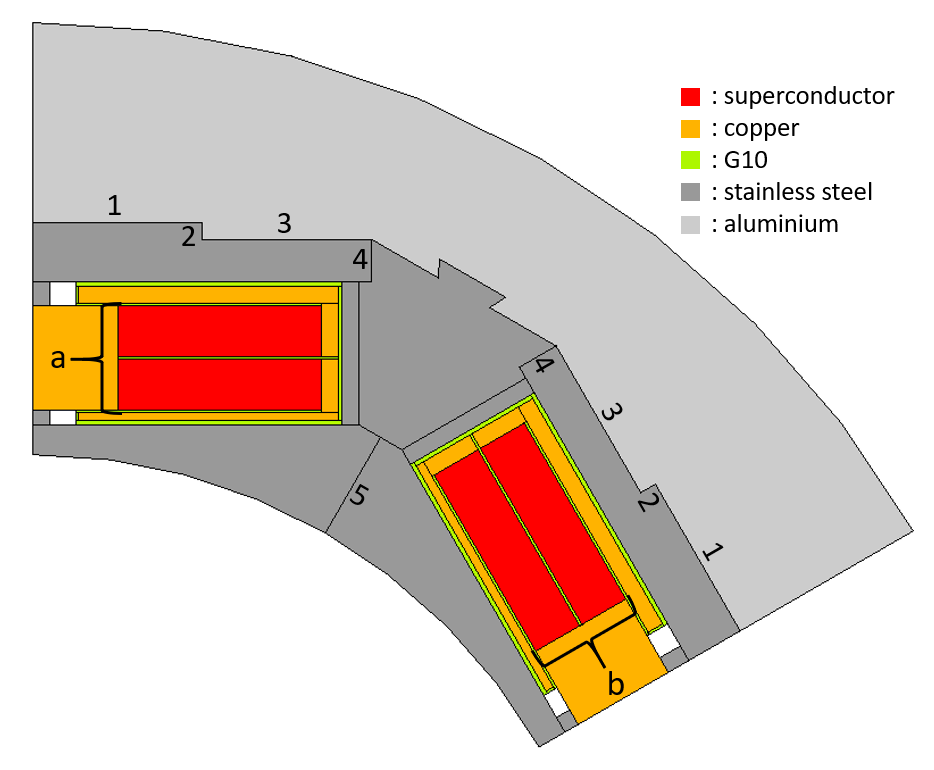} \\
    \includegraphics[width=\columnwidth]{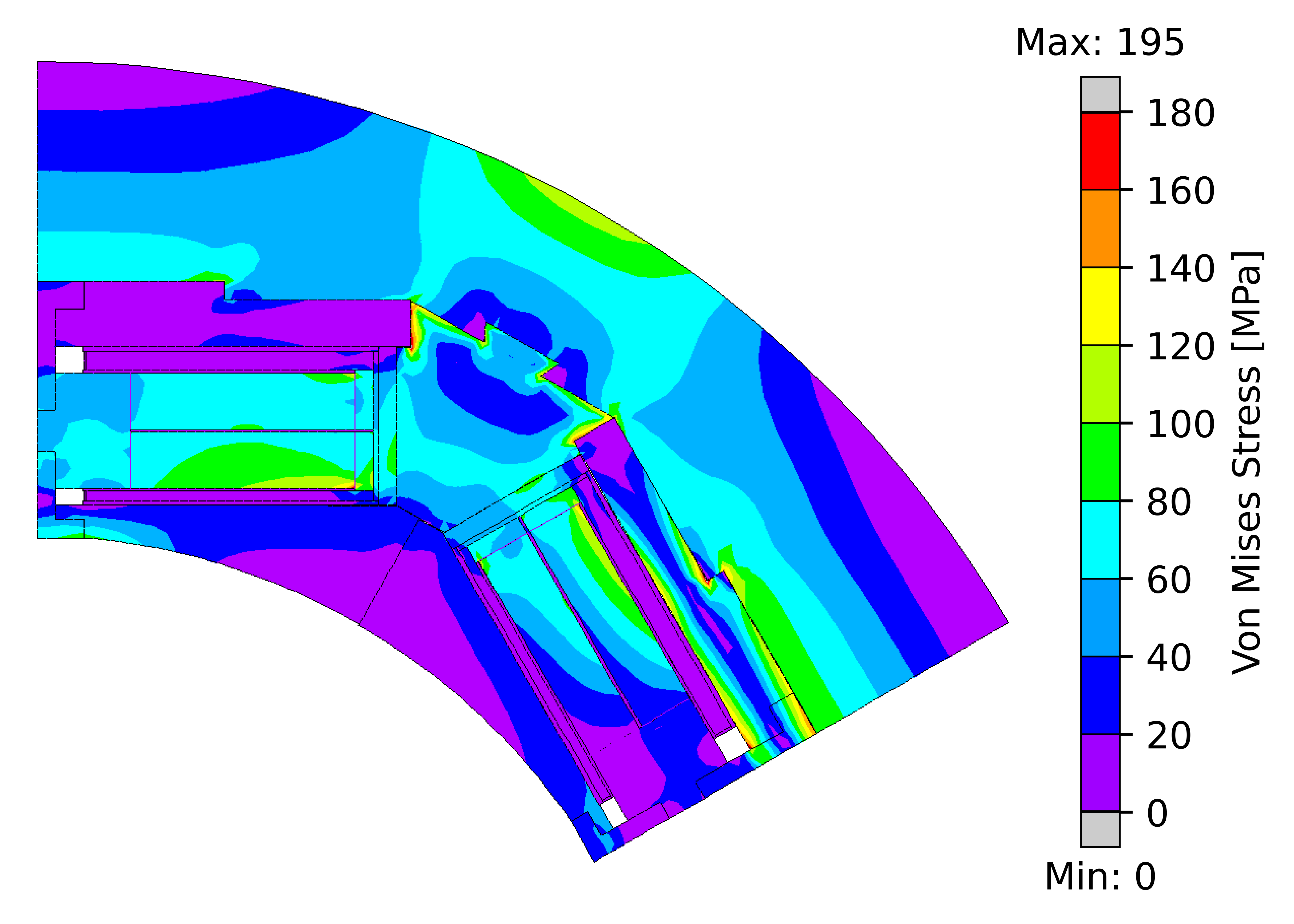}
    \caption{Upper plot: CAD cross-section of a sixth of the TOUHTATIS sextupoles. Lines and gaps denoted 1–5 are design variables to be optimised. Lower plot: von Mises stress on this cross-section.}
    \label{fig:touhtatis}
\end{figure}

The \alesia{} platform was firstly used for the superconducting magnet of the 28 GHz ASTERICS ion source \cite{Simon2023} developped for the NEWGAIN project and that will be installed on the SPIRAL2 Linac at GANIL in Caen (France). For this magnet who have been designed at CEA and is now under construction in the industry, a parametric sensitivity analysis of component positioning errors was conducted using \textsc{opera 3d} thought \alesia{}. The objective was to quantify the impact of geometric perturbations (misalignments and displacements) on the magnetic field topology and confinement properties. 
This study aims at defining acceptable assembly tolerances and identifying the most critical components.

Building upon ASTERICS, CEA is investigating the design of a next-generation 45 GHz ECRIS, named TOUHTATIS, using HTS conductor, targeting higher ion density and improved mass-over-charge ratio.
The magnetic configuration follows the classical ECR topology: a sextupole composed of six racetrack double-pancake coils, surrounded by three solenoids (injection, middle and extraction).
However, due to the rapid decline of LTS critical current under high magnetic fields, TOUHTATIS relies on REBCO HTS tapes to achieve the required magnetic field levels.
The use of HTS enables increased compactness and operation at 20 K instead of 4 K, which improve energy efficiency, and reduced operational costs and maintenance.
The magnetic design was developed using \textsc{rat}, while mechanical behaviour was assessed using \textsc{ansys}.

\begin{figure}
    \centering
    \includegraphics[width=\columnwidth]{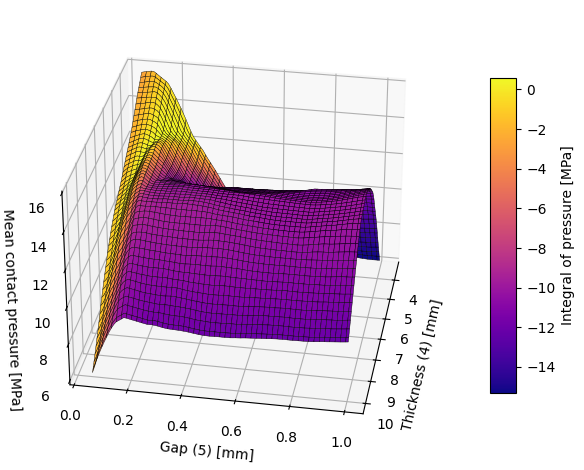}
    \caption{Mean contact pressure and pressure integral-based metric (evaluated along b in Fig.~\ref{fig:touhtatis}) from the thickness of the upper steel plate (called 4 in Fig.~\ref{fig:touhtatis}), and the gap between lower steel plates (called 5 in Fig.~\ref{fig:touhtatis}). Higher contact pressure and integral of pressure are better.}
    \label{fig:touhtatis2}
\end{figure}

As part of the TOUHTATIS effort, and to demonstrate the viability of using HTS for a 45 GHz ion source, CEA is currently developing a short-model design for an 18 GHz ECRIS that will serve as a proof of concept. 
It is to optimise the design of this short model that \alesia{} have been used.
A key design objective of TOUHTATIS is the modularity of its sextupole racetracks. 
Each racetrack assembly consists of REBCO tapes, copper stabiliser, and fiberglass electrical insulation (G10). They are inserted into a stainless steel casing, towards an aluminum support ring with stainless steel teeth.
The mechanical performance of this assembly critically depends on contact pressure distribution between components.
Figure \ref{fig:touhtatis} presents the sextupole CAD configuration.
Several geometric parameters, in particular the lines and gaps denoted 1–5 in Fig.~\ref{fig:touhtatis}, were identified as design variables. 
The optimisation objective was to inmprove the mean and standard deviation of the contact pressure on a and b.
Two complementary strategies were implemented within \alesia{} and based on mechanical simulations from \textsc{ansys}.
A direct optimisation using evolutionary computation (similar to Section~\ref{s:nb3sn} or Ref.~\cite{Minenna2026}), whose goal is to converge towards the optimal values of the contact pressures, allowing to quickly find a suitable solution.
In addition, supervised surrogate models were trained based on 1400 random simulations.
Those models included a Random Forest, a Gaussian process, and a Neural Network.
These models enabled a rapid multi-dimensional response surface exploration, an identification of dominant parameters, a correlation quantification, and a better understanding of the design trade-offs.
Figure~\ref{fig:touhtatis2} presents a predictive model from a fully connected deep neural network using a pipeline, combining feature scaling and neural network modelling with four hidden layers using the ReLU (Rectified Linear Unit) activation function and the Adam optimiser.

\subsection{Extraction Systems}

\begin{figure}
    \centering
    \includegraphics[width=\columnwidth]{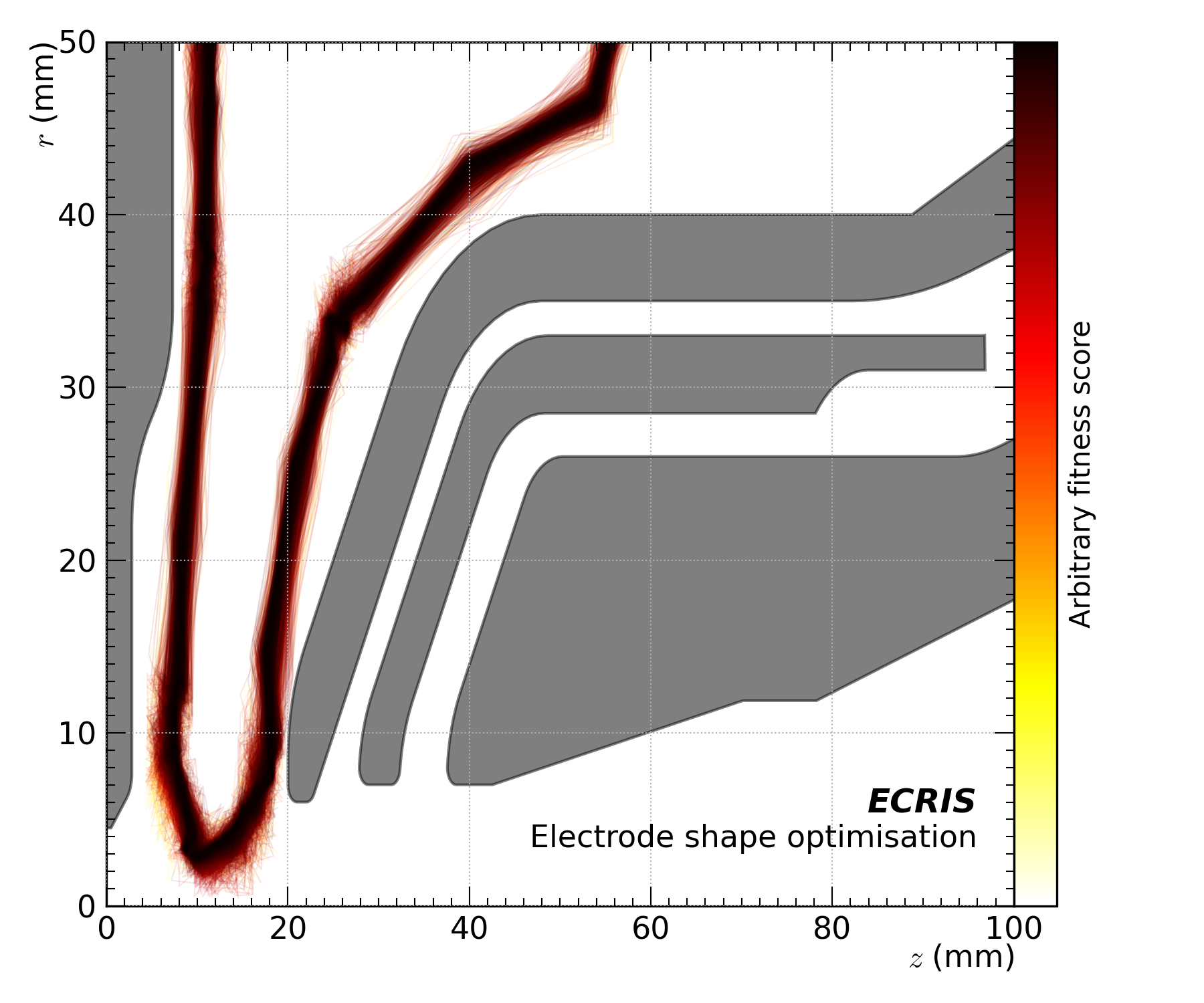}
    \caption{Superposition of 27225 valid polygons, composed of 35 points, shaping the main repeller electrode of an ECRIS extraction system. The fitness score comprises the beam emittance, beam current and validity of the shape. The grey regions are other electrodes with the aims to optimise them as well.}
    \label{fig:ecris}
\end{figure}

The \alesia{} platform is also employed \cite{Bolzon2025,Tuske2025} for the optimisation of ECRIS beam extraction simulations. It has been integrated with the \textsc{ibsimu} software for beam dynamics. 
\alesia{} can adjust the ion source parameters used as inputs for \textsc{ibsimu} simulations (e.g., gap length, electrode aperture hole diameter, and current density at the plasma electrode aperture) using evolutionary algorithms, such as a particle swarm algorithm, to rapidly enhance the extraction design for an optimised beam with low emittance.
Our goal is to automate the design of the electrodes and potentials through topological optimisation (see Fig.~\ref{fig:ecris}) . \alesia{} can manage shape variations of electrodes, initiate \textsc{ibsimu} simulations, retrieve the beam emittance from \textsc{ibsimu}, update the electrode shape using active learning methods, and iteratively refine the shape of electrodes until the specified tolerances are met. 
Currently, a promising method combining deep reinforcement learning with evolutionary optimisation strategies is under investigation to propose physically valid electrode shapes.

\section{Quench Protection}
\label{s:quench}

A quench is the rapid transition of a superconducting material to its normal (resistive) state, that may lead to the release of the very large stored energy into a small volume of conductor. 
It can cause irreversible damage to magnets and must be mitigated.

\subsection{Anomaly Detection}

To prevent quench damage on operating magnets, it is crucial to detect quickly any anomalies and activate protective measures.
The quench detection is particularly challenging for recent HTS due to their very slow normal-zone propagation velocities. 
To address this, CEA is developing digital twins of the superconducting systems. 
Currently, these models are trained using voltage data from REBCO pancake coils, leveraging in-house Partial Element Equivalent Circuit (PEEC) codes \cite{Genot2022}. 
These models integrate both classifiers and regressors. 
The classifiers (mostly decision trees and random decision forests) assess detected anomalies to determine whether they will lead to a rapid quench, a slow quench, or if the system is in a recovery state. 
Meanwhile, the regressors (convolutional neural networks and long short-term memory) forecast the temporal evolution of voltage and magnetic fields within the coils. 
Additionally, models are developed at various time scales: fast models analyse a few milliseconds of input data, while more accurate models use tens of milliseconds of input data. 
The long term goal is to enhance these models by integrating experimental data to improve their accuracy and reliability.

\subsection{Propagation Velocities}

Accurately estimating the longitudinal and transverse propagation velocities, influenced by the conductor properties and insulation layers, in the normal zone during a quench event is a critical challenge. 
These velocities govern how rapidly this zone expands and influence the temperature map, hence impacting the protection strategy and overall system safety.
At present, the most reliable method to evaluate these propagation velocities involves magnetoresistive time-dependent finite element simulations, used at CEA \cite{Juster2004,Juster2010}. 
While these multiphysics models provide high-fidelity results, they are computationally expensive and require substantial modelling effort limiting their practical use. 
CEA is currently investigating surrogate models aimed at providing fast and sufficiently accurate estimations of the propagation velocities.
The resulting propagation velocities will be integrated into a fast 3D adiabatic model of the quench called \textsc{qaesar} (Quench quasi-Analytic Equivalent Solver in Adiabatic Regime).
This will enable rapid parametric analyses, to improve design guidelines, and more efficient protection strategies, while, in principle, maintaining a strong physical foundation.

\section{Conclusion}
\label{s:conclu}

Future generations of superconducting magnets for particles accelerators, including Nb-Ti, Nb$_3$Sn or HTS magnets, can greatly benefit from the emergence of artificial intelligence tools and advanced optimisation methods.

This paper presents several ongoing case studies at CEA illustrating how data-driven approaches, most of them implemented within the \alesia{} platform, can support complex multiphysics design and accelerate optimisation workflows. 
The platform will be further developed to integrate additional physics software tools and new advanced optimisation methods.

Beyond particle accelerators, \alesia{} is being applied to a broad range of projects including MRI magnets and magnets for fusion.In parallel, CEA is exploring new optimisation strategies, including physics-informed neural networks, as well as the use of generative AI to automatically construct optimisation workflows based on knowledge extracted from previous projects.

\appendix

\section*{Acknowledgements}
The authors thank the members of DACM at CEA, IRFU for fruitful discussions.

\section*{Author contributions}
This manuscript brings together several projects supported over the years by many members of DACM at CEA, IRFU. 
The lead instigator of this work is Damien F.\ G.\ Minenna.
This manuscript was mainly written by Damien F.\ G.\ Minenna, Guillaume Dilasser, Thibault de Chabannes and Thibault Lecrevisse, and used data and plots from Thomas Achard and Jason Le Coz.
Damien F.\ G.\ Minenna, Robin Penavaire and Valerio Calvelli are the main contributors to the ALESIA platform, which was used on several aspects of this work.  
All authors have read and approved the final manuscript.

\section*{Data availability}
The data that support the findings of this study are available from the corresponding author upon reasonable request.

\section*{Competing interests}

The authors declare no Conflict of interest.

\bibliography{Optimisation}

\end{document}